\documentclass[preprint,preprintnumbers,amsmath,amssymb]{revtex4}
\usepackage{times}
\usepackage{graphicx}
\usepackage{dcolumn}
\usepackage{bm}

\begin{document}

\title{Redundancies in Large-scale Protein Interaction Networks}
\author{Manoj Pratim Samanta and Shoudan Liang}
\affiliation{NASA Advanced Supercomputing Division, NASA Ames Research Center, Moffet Field, CA, USA}
\date{\today}
\vspace{-0.3in}

\begin{abstract}
Understanding functional associations among genes
discovered in sequencing projects is a key issue in
post-genomic biology~\cite{Gavin,Ho,Ito,Uetz,Tong1,Zhu1,Grunenfelder}.
However, reliable interpretation of the protein interaction data has been
difficult~\cite{Uetz2,Mering-short}. In this work, we show that if two proteins
share significantly larger number of common interaction partners than random,
they have close functional associations. Analysis of publicly available data from 
{\it Saccharomyces cerevisiae} reveals more than 2800
reliable functional associations, 29\% of which involve at least one 
unannotated protein. By further analyzing these associations, 
we derive tentative functions for 81 unannotated proteins
with high certainty.
\end{abstract} 


\maketitle

A large number of genes discovered in sequencing projects remain functionally
unannotated, motivating significant research in post-genomic biology. High-throughput 
experiments
such as genome-wide monitoring of mRNA expressions as well as 
protein-protein interaction networks are
expected to be fertile sources of information to derive their
functions~\cite{Gavin,Ho,Ito,Uetz,Tong1,Zhu1,Grunenfelder}.
However, a high rate of false positives~\cite{Uetz2,Mering-short} as well as the 
sheer volume of the data are making reliable interpretation of these experiments
difficult.  

In this work, we are able to overcome these difficulties by using a 
statistical method that forms reliable functional associations between proteins from 
noisy genome-wide interaction data. Our method ranks the statistical
significance of forming shared partnerships for all protein pairs in the interaction
network and shows that if two proteins share significantly larger number of common
partners than random, they have close functional associations.
 In the supplement, we derive more than 2800 pairs of high quality associations for 
{\it S. cerevisiae} involving 852 proteins. The method is not overly sensitive from 
the false positives widely present in the two-hybrid
data. Even after adding 50\% randomly generated interactions to the 
measured dataset, we are able to recover almost all ($\sim 90\%$) of the original associations.
The modular nature of the interaction network~\cite{Hartwell} is 
revealed by the clustering of these associations. From the derived modules, we are
able to predict functions for 81 unannotated proteins with high certainty. It has been an 
encouraging sign that the functions of some of these proteins were recently 
annotated by the SGD database~\cite{SGD-short} from other sources after the completion of our
work, and all but one (22 out of 23) of our predictions proved to be correct.

Our strategy of assigning statistical significance is to compare the measured
protein interaction network with a random network of the same 
size~\cite{Jeong,Jeong2,Maslov}.
The deviation of the measured network from randomness is presumed to
reflect its biological significance. Non-random nature of the large-scale protein
interaction network has been discussed in earlier work~\cite{Jeong,Maslov,Bader}.
In one example, it was 
observed that the connectivities of the proteins in the measured interaction networks 
closely followed a power-law distribution instead of the exponential distribution 
expected from random networks~\cite{Jeong,Mering-short,Maslov,Bader}. 
Useful biological prediction regarding the lethality of the null mutants 
lacking those highly connected proteins 
could be made from such non-random behavior~\cite{Jeong}.

We hypothesize that if two proteins
have significantly larger number of common interaction partners in the measured 
data-set than what is expected from a random network, it would suggest close 
functional links between them. To validate this hypothesis, we rank all possible protein
pairs in the order of their probabilities~\cite{note_equation}
 for having the experimentally measured
number of common interaction partners.
If the computed probability is extremely small, it signifies that the chosen protein pair has
an unusually large number of common 
partners. Such pairs are considered for further analysis, as we discuss in the paper.

\begin{figure}
\centering
\includegraphics[width=8cm]{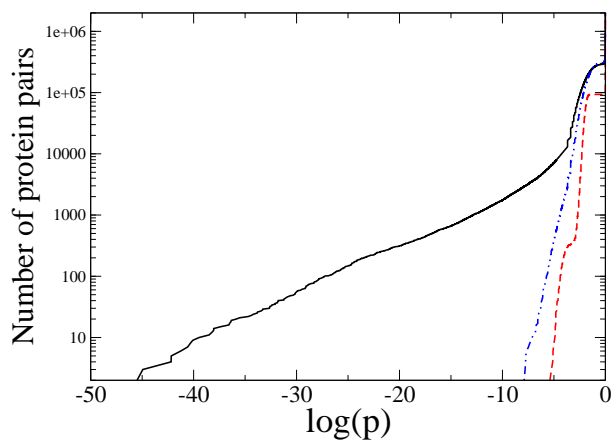}
\caption{Probabilities of associations for all possible protein pairs derived using our 
method. Solid black line: measured protein interaction network~\cite{Deane};
broken red line: a random network of similar size constructed by connecting randomly 
chosen nodes; dotted green line: a random network constructed from the measured
network keeping its power-law connectivity property unchanged~\cite{Maslov}. 
The probabilities of associations for the measured network are up to
40 orders of magnitude lower than the random networks.}
\label{fig:probability}
\end{figure}

The described method is applied on the available experimental data from budding yeast 
({\it S. cerevisiae}) collected in the DIP database from several 
sources~\cite{note_DIP,Deane}. In Fig.~\ref{fig:probability}, we show a 
plot with probabilities for all protein pairs in the network
sorted in increasing order. For comparison, we also show corresponding probabilities 
for a random network of similar size, as well as a randomized version of
the measured network. The random network has the same number of nodes and edges as the
measured network, but the connections are made from a uniform distribution.
The randomization of the experimental network is done using a method similar to 
Ref.~\cite{Maslov}. The method allows us to maintain the power-law nature of the network.
As we observe from the plot, the probabilities of some of the associations in the 
measured network are up to 40 orders of magnitude lower than both of the randomly
constructed networks. Therefore, it is safe to conclude that those associations are
not artifacts due to experimental noise, but contain biologically meaningful
information.  It is also clear from the plot that such low probability associations
did not arise from the scale-free nature of the network~\cite{Jeong}.

To understand what biological information is provided by such low-probability
pairs, we inspect all pairs with probabilities below a cutoff value of 
$10^{-8}$~\cite{note_cutoff}.  The detailed list is provided as a 
supplement~\cite{note_supplement} as well as from our website~\cite{note_web}.
The group consists of 2833 protein pairs involving 852
proteins.  A strong functional link is observed among proteins in these pairs thus
validating our hypothesis.  This is illustrated in 
Table~\ref{table:prots}, where we present the ten pairs with the lowest probabilities.
As we can see from the table, both proteins usually either belong to the same complex 
or are parts of the same functional 
pathway. Same trend is generally true for the larger dataset presented in the 
supplement. By manually inspecting the top 100 pairs, we found that in over 95\%
of them both proteins have similar function.

We can take advantage of the above observation to predict the functions of 
the unannotated proteins. About 29\% of the 2833 chosen pairs 
contain at least one unannotated protein~\cite{note_Mering}. To assign a function
to any one of them, we determine the other proteins with which it forms associations.
As an example, in Table~\ref{table:prots2}
we show that the unannotated protein $YKL059C$ shares partners with many proteins
involved in transcription. Therefore, it is most likely also involved in transcription.
Moreover, from the low probabilities of associations with CFT2 and CFT1,
we strongly suspect that that YKL059C is involved in pre-mRNA 3' end processing. 
This is further confirmed by the clustering method that we present below.
Our website provides an interactive tool for users to search for the close associates
of any query protein and thus derive its putative function~\cite{note_web}.

\begin{table}
\begin{tabular}{|l|l|l|l|}
\hline
Protein 1 & Protein 2 & Log(p) & Function \\
\hline
MYO3 & MYO5 & -47.41 & Class I myosins \\
ROX3 & SRB6 & -46.12 & Mediator complex \\
KRR1 & PWP2 & -45.50 & snoRNA complex \\
ROX3 & MED2 & -44.94 & Mediator complex \\
MED2 & SRB6 & -42.19 & Mediator complex \\
ATP1 & ATP2 & -42.17 & ATP complex \\
KAP95 & SRP1 & -41.25 & Protein import-export\\
PRE1 & RPN10 & -40.58 & Spliceosome complex\\
YKR081C & YNL110C & -40.33 & Both unannotated\\
RPT1 & RPN6 & -40.07 & Spliceosome complex\\
\hline
\end{tabular}
\caption{The ten protein pairs with the lowest probabilities~\cite{note_equation}
 based on our method, along with
their functions. We find both of the proteins in these
pairs to belong to either the same complexes or the same functional pathways. 
The complete list is provided as a supplement.
}
\label{table:prots}
\end{table}

\squeezetable

\begin{table}
\begin{tabular}{|l|c|}
\hline
Associations of YKL059C & Log(p)\\
\hline
CFT2[T] & -32.430607 \\
CFT1[T] & -30.151475 \\
YSH1[T] & -28.320081 \\
PTA1[T] & -27.843331 \\
PAP1[T] & -27.410048 \\
REF2[T] & -25.048611 \\
PFS2[T] & -24.638901 \\
YTH1[T] & -23.247919 \\
FIP1[T] & -21.609526 \\
HCA4[T] & -21.285573 \\
YGR156W[U] & -17.961537 \\
RNA14[T] & -17.732432 \\
SWD2[U] & -14.407007 \\
GLC7[C] & -13.284243 \\
YOR179C[T] & -12.636400 \\
PCF11[T] & -8.857110 \\
\hline
\end{tabular}
\caption{ Categories - T: transcription, U: unannotated protein, 
C: cellular fate/organization. Most of the associations of YKL059C are involved in 
transcription and therefore it is also expected to do the same.  From its very low 
probabilities~\cite{note_equation} of associating with CFT1 and
CFT2, it is strongly suspected to be involved in pre-mRNA 3' end processing.
Our website provides an interactive tool to search for the associates of any 
protein~\cite{note_web}.
}
\label{table:prots2}
\end{table}

\begin{figure*}
\centering
\includegraphics[width=12cm]{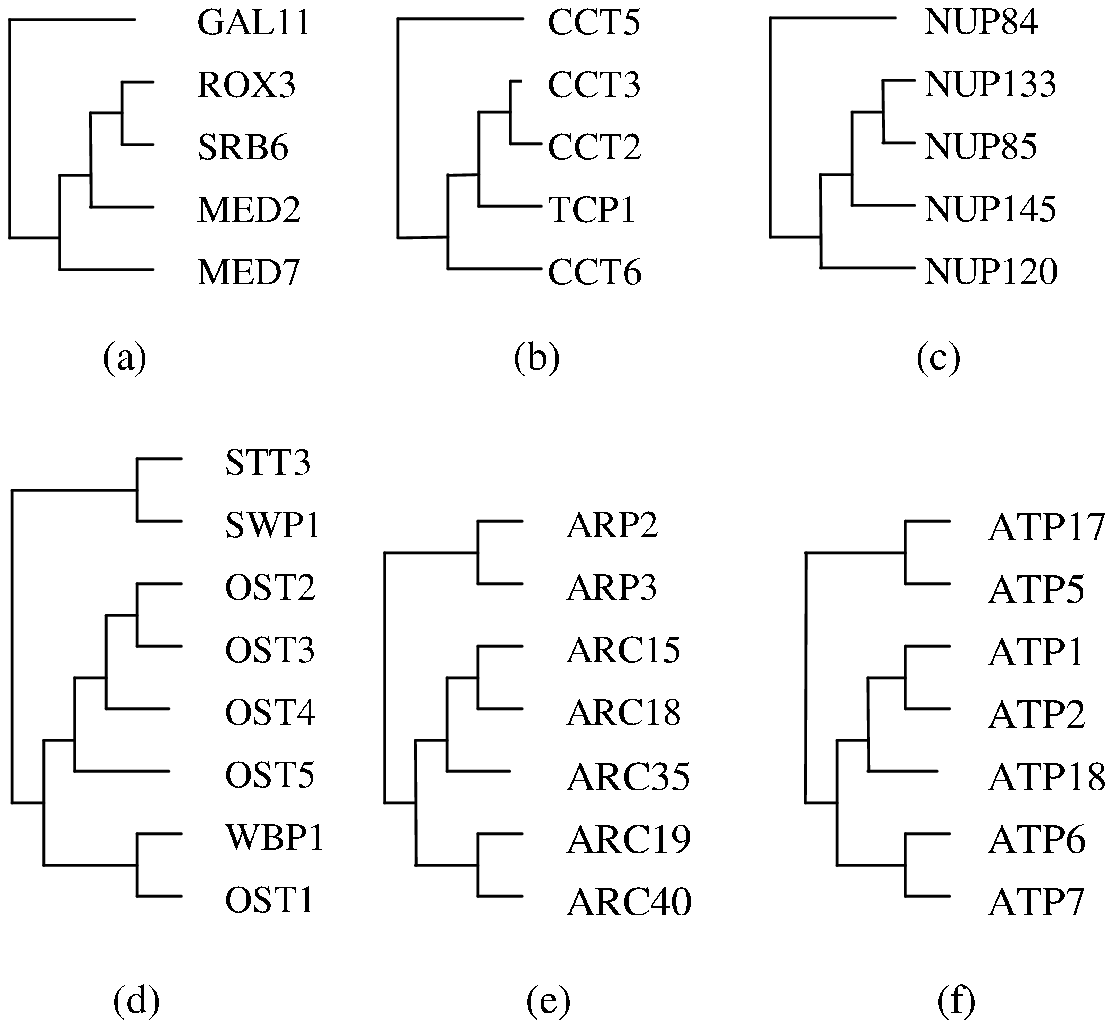}
\caption{Functional modules obtained by clustering the low-probability associations
using an algorithm described in our paper. All proteins from each of these
derived modules belong to same functional complexes.     (a) Pol II 
transcription mediator complex, (b) chaperon ring complex, (c) nuclear pore complex, 
(d) oligosaccharyl transferase complex,  (e) Arp2/3 complex, (f) ATP synthase
complex. The complete list of modules is provided in the supplemental table.}
\label{fig:cluster}
\end{figure*}

Since functionally related proteins form strong associations with each other,
this can be used as the basis for an algorithm to cluster them into functional modules.
We derive $202$ modules [Fig.~\ref{fig:cluster}]
 from the associations and then compare the annotations
of constituent proteins.
$163$ of the derived modules have all proteins annotated in the SGD
database~\cite{SGD-short} and we find 149 of them (about $92\%$) to have all members of 
the module from the same functional
complex or pathway.  Therefore, if an unannotated protein belongs to the same
modules with other proteins of known functions, we can predict its functions to be
the same as the other ones with high confidence.
By analyzing the derived modules, we predict functions for $81$ 
unannotated proteins and  present them in Table~\ref{table:function}.

\begin{figure}
\centering
\includegraphics[width=3.7cm]{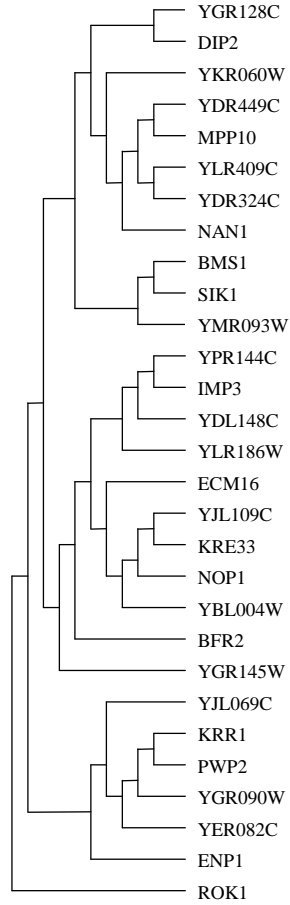}
\caption{A module identified by our method consisting of proteins presumably
involved in assembly and maintenance of small nucleolar ribosomal complex.}
\label{fig:one}
\end{figure}

\begin{figure}
\centering
\includegraphics[width=3.5cm]{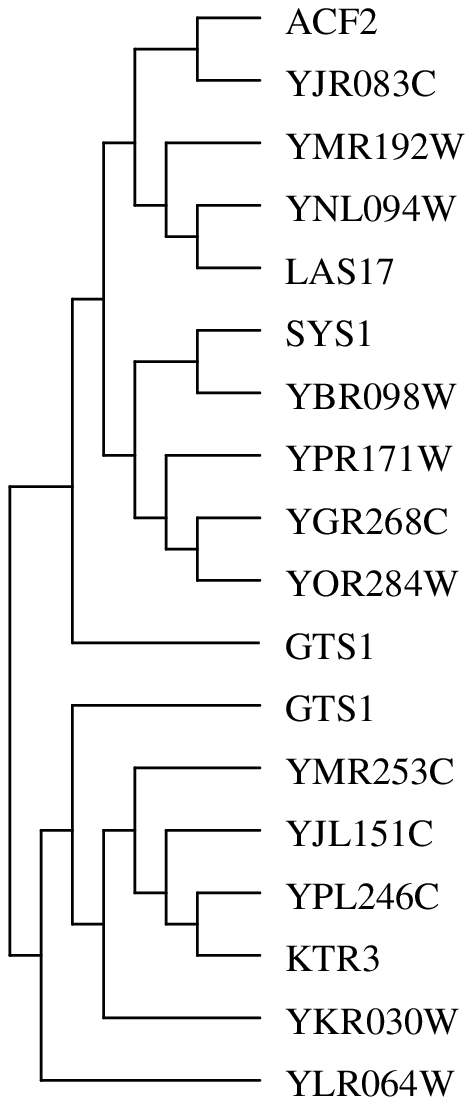}
\caption{A module identified by our method consisting of proteins presumably involved
in actin cytoskeleton organization and protein vacuolar transport.}
\label{fig:two}
\end{figure}

We note that the chosen cut-off value ($10^{-8}$) is not a sharp threshold. 
As the number is increased, the amount of biologically
meaningful information degrades gradually. In the case of 
the modules, their numbers and sizes increase with increasing cut-off. As an 
example, for the well-studied mediator complex shown in figure~\ref{fig:cluster}(a), 
as we increase the cut-off value, more proteins known to be part of the complex come 
together. We find that even with cut-off as high as $2\times10^{-4}$, the proteins
included in the mediator module are genuinely related to the complex. In our website
we present an interactive program that allows users to choose different cut-off values 
and obtain the corresponding modules. Among the additional modules derived with
higher threshold, we find two that contain mostly unannotated proteins and therefore
are possibly large complexes not yet well studied by experimentalists. 
One of them is suspected to be involved in actin cytoskeleton organization and 
protein vacuolar targeting and the other one in splicing, rRNA processing and
snoRNA processing. We present them in Figs.~\ref{fig:one} and \ref{fig:two}
expecting their identification to spur additional interest among yeast biologists.

The method presented here has several advantages.
Firstly, it is not sensitive to random false positives.  To illustrate, we added
connections randomly increasing the average number of interactions by 50\%
and were still able to recover 90\% of the top 2833 associations.
Secondly, the method is not biased by the number of partners a protein has.
As an example, JSN1, a nuclear pore protein, has the largest number
of interactions in the measured dataset, but none of the 2833 associations derived by
our method contains JSN1. Among the drawbacks, our method cannot extract much 
information about proteins with none or very few interactions in the dataset.

In conclusion, we derived functional modules and reliably predicted
functions of unannotated proteins from the existence of abnormally large number
of shared interaction partners in the protein-protein interaction network.
We believe the real power of the method will be in studying the higher eukaryotes,
where higher fraction of genes has unknown functions.
Moreover, the method is applicable to other forms of networks, such as the Internet,
metabolic networks, social networks and predator-prey networks.

\bibliographystyle{unsrt}
\bibliography{main}

\begin{thebibliography}{10}
\bibitem{note_Mering}
Here, we use the functional classes and annotations provided in
  Ref.~\cite{Mering-short}. However, SGD database is constantly updating the
  annotations for the proteins as new information becomes available. Therefore,
  the actual number of unannotated proteins is lower than this source.
\bibitem{note_SGD}
These proteins recently got annotated in the SGD database. Except for LSG1, all
  other predictions proved to be correct.
\end{thebibliography}

\squeezetable
\topmargin -.5in
\linespread{1}
\begin{table*}
\begin{tabular}{|c|l|}
\hline
Protein & Predicted Function \\
\hline
YFR024C (YSC85), BZZ1~\cite{note_SGD}, YNL094W (APP1), YMR192W (APP2) & Actin filament organization \\
\hline
YGR268C (HUA1), YOR284W (HUA2), YPR171W (BSP1) & Actin patch assembly\\
\hline
YJR083C (ACF4) & Actin cytoskeleton organization and biogenesis \\
\hline
YDR036C (MRP5) &  Protein biosynthesis in mitochondrial small ribosomal subunit\\
\hline
YKL214C (YRA2)~\cite{note_SGD} & mRNA processing/RNA metabolism \\
\hline
YNL207W (RIO2) & Nucleolar protein involved in 40S ribosomal biogenesis \\
\hline
\begin{minipage}{3 in}
YLR409C (UTP21), YKR060W (UTP30), YGR090W(UTP22), YER082C(UTP7)~\cite{note_SGD}, YJL069C(UTP18)~\cite{note_SGD}, ENP1
\end{minipage}
 & Associated with U3 snoRNA and 20S rRNA biosynthesis\\
\hline
YMR288W (HSH155)~\cite{note_SGD} & snRNA binding involved in mRNA splicing \\
\hline
YHR197W (IPI2), YNL182C (IPI3), YLR106C (MDN1)~\cite{note_SGD} & Ribosomal large subunit assembly and maintenance \\
\hline
YGR128C (UTP8)~\cite{note_SGD} & Processing of 20S pre-rRNA \\
\hline
YGR215W (RSM27)~\cite{note_SGD}, YGL129C (RSM13)~\cite{note_SGD}& 
Structural constituent of ribosome \\
\hline
YDL213C (NOP6) & rRNA processing/transcription elongation \\
\hline
YNL306W (MRPS18)~\cite{note_SGD} & Mitochondrial small ribosomal subunit \\
\hline
\begin{minipage}{3 in}
YPR144C (UTP19), YDL148C (NOP14)~\cite{note_SGD}, 
YLR186W (EMG1), YJL109C (UTP10)~\cite{note_SGD}, 
YBL004W (UTP20) 
\end{minipage}
& snoRNA binding, 35S primary transcript processing \\
\hline
YGL099W (LSG1)~\cite{note_SGD}, YDR101C (ARX1)& 27S pre-rRNA ribosomal subunit \\
\hline
BRX1, YOR206W (NOC2), FPR1	& Biogenesis and transport of ribosome \\
\hline
YOR145C (DIM2) & 35S Primary transcript processing and rRNA modification \\
\hline
YEL015W (DCP3) & 
Deadenylation dependent decapping and mRNA catabolism \\
\hline
NHP10, RFX1~\cite{note_SGD} & 
Modification of chromatin architecture/transcription\\
\hline
YDR469W (SDC1)~\cite{note_SGD} & Chromatin silencing and histone methylation \\
\hline
YPL070W (MUK1) & 
Transcription factor (or its carrier) \\
\hline
YLR427W (MAG2) & DNA N-glycosylase involved in DNA dealkylation \\
\hline
YDL076C (RXT3), YIL112W (HOS4) & Histone deacetylase complex involved in chromatin silencing \\
\hline
IST1 &	Trancription initiation factor\\
\hline
HCR1~\cite{note_SGD}	& Translation initiation as part of eIF3 complex \\
\hline
YDL074C (BRE1) & Chromosome condensation and segregation process \\
\hline
YGR156W (PTT1)~\cite{note_SGD}, YKL059C (MPE1)~\cite{note_SGD} & 
mRNA cleavage and polyadenylation specificity factor\\
\hline
YGR089W (NNF2) & Chromosome segregation (spindle pole) and mitosis \\
\hline
YGL161C(YIP5) ,YGL198W (YIP4) & Vescicle mediated transport \\
\hline
\begin{minipage}{3 in}
YPL246C (QUT1), YJL151C (SNA3), YGL104C (VPS73)~\cite{note_SGD}, YKR030W (MSG1)
\end{minipage}
& Cell wall synthesis / protein-vacuolar targeting \\
\hline
YBR098W (MMS4) & Golgi to endosome transport and vescicle organization \\
\hline
YHR105W (YPT35) & Golgi to vacuolar transport \\
\hline
YBL049W (MOH1), YCL039W (MOH2) & Both same function. Possibly linked with vacuolar transport\\
\hline
YDL246C (SOR2) & 
Possibly involved in fructose and mannose metabolism\\
\hline
YMR322C (SNO4) & Pyridoxine metabolism \\
\hline
YDR430C (CYM1) & Protein involved in pyurvate metabolism \\
\hline
YJL199C (MBB1), YPL004C (LSP1), YGR086C (PIL1) & Metabolic protein \\
\hline
YLR097C (HRT3) & Nuclear ubiquitine ligase \\
\hline
YKR046C (PET10) & ATP/ADP exchange \\
\hline
YEL017W (GTT3) & Protein linked with glutathione metabolism \\
\hline
ITC1	& Chromatin remodeling \\
\hline
YGR161C (RTS3) & Protein phosphatase 2A complex \\
\hline
EFD1 & DNA replication and repair \\
\hline
YML117W (NAB6) & Nuclear RNA binding \\
\hline
YLR432W (IMD3) & RNA helicase involved in mRNA splicing \\
\hline
YJU2, YGR278W (CWC22), YDL209C (CWC2)~\cite{note_SGD} & 
Spliceosome complex involved in mRNA splicing \\
\hline
\begin{minipage}{3 in}
YGR232W (NAS6)~\cite{note_SGD}, YGL004C (RPN14)~\cite{note_SGD}, 
YLR421C (RPN13)~\cite{note_SGD} 
\end{minipage}
& Proteasome complex \\
\hline
\end{tabular}
\caption{
Predicted functions of previously unannotated proteins.
}
\label{table:function}
\end{table*}

\thispagestyle{empty}

\appendix

\section{Supplemental Informations (Theory)}
\subsection{Methods}

\begin{figure}
\centering
\includegraphics[width=6cm]{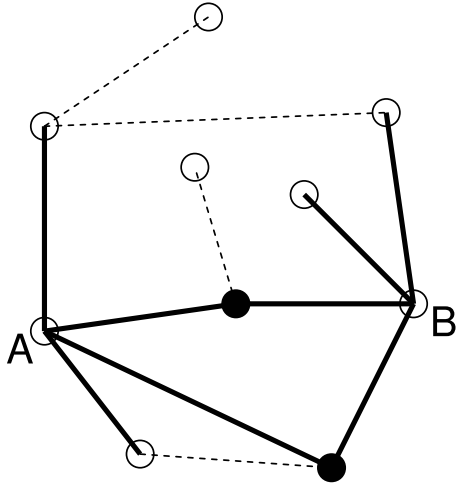}
\caption{In the above interaction network, both proteins A and B have 4 partners
($n_1$ and $n_2$).
Two of the partners (marked by dark circles) are shared by both of them.
We compute the probability for such an event to occur in a
random network. If the computed probability is low, perhaps two proteins are
redundant in their functions.
}
\label{fig:schematic}
\end{figure}

\subsubsection{Mathematical Expression for Probability}
In a network of $N$ proteins, the probability that two proteins with $n_1$ and
$n_2$ partners [Fig.~\ref{fig:schematic}] share $m$ common partners is given by
\begin{equation}
P(N,n1,n2,m)=
\frac{
\left( \begin{array}{c} n_1\\ m \end{array} \right)
\left( \begin{array}{c} N - n_1\\ n_2 - m \end{array} \right)
}
{
\left( \begin{array}{c} N\\ n_2 \end{array} \right)
}
=\frac{
(N-n_1)! (N-n_2)! n_1! n_2!
}
{
N! m! (n_1-m)! (n_2-m)! (N-n_1-n_2+m)!
}
\label{eq:P}
\end{equation}
The above expression is symmetric with respect to interchange of $n_1$ and $n_2$.
Eq.~\ref{eq:P} is derived in the following manner.
It is a ratio where the denominator is the  total number of ways two proteins can
have $n_1$ and $n_2$ partners given by
\begin{equation}
\left( \begin{array}{c} N\\ n_1 \end{array} \right)
\left( \begin{array}{c} N\\ n_2 \end{array} \right),
\end{equation}
whereas the numerator is the number of cases among them
where $m$ of those partners are common to both of them. It is expressed as
\begin{equation}
\left( \begin{array}{c} N\\ m \end{array} \right)
\left( \begin{array}{c} N-m\\ n_1-m \end{array} \right)
\left( \begin{array}{c} N - n_1\\ n_2 - m \end{array} \right)
=
\left( \begin{array}{c} N\\ n_1 \end{array} \right)
\left( \begin{array}{c} n_1\\ m \end{array} \right)
\left( \begin{array}{c} N - n_1\\ n_2 - m \end{array} \right).
\label{eq:F}
\end{equation}
The numerator can be derived using the following argument. In the combinatorial
product on the left hand side, the first term represents the number of
ways $m$ common partners can be chosen from all $N$ proteins.
For the first protein, we choose $n_1-m$ remaining
partners out of remaining $N-m$ proteins. This is the second term in the
product. Subsequently for the second protein, we choose $n_2-m$ remaining
partners none of which match any of $n_1$ partners of the first
protein. This contributes to the third term.

For the calculations in our paper, the results are approximately the same, whether we
compute the probabilities for pairs with exactly $m$ common partners  or we
compute for $m$ or more partners. It can be checked from the expression of
probability in Eq.~\ref{eq:P}, that probability terms for increasing $m$
fall inversely with $N$. Since $N$ for our case is about $5000$, the
additional terms in the probability expression are negligible.

\begin{figure}
\includegraphics[width=8cm]{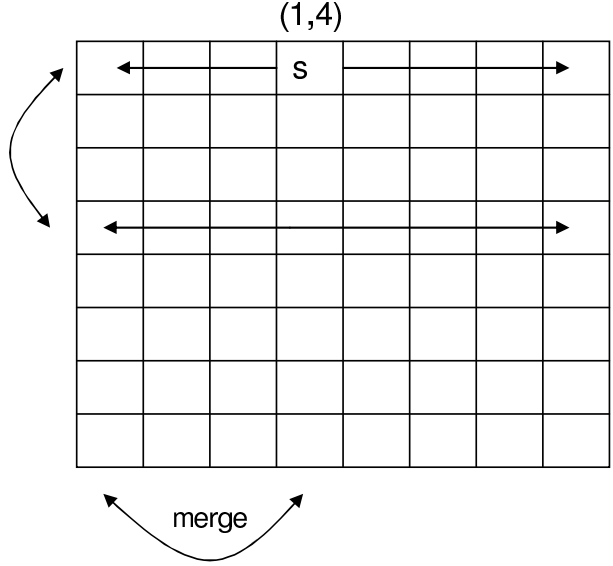}
\caption{In our clustering algorithm, we start with a matrix with $p$-values
for all pairs. If the element $(m,n)$ has the lowest $p$-value, a cluster is
formed with proteins $m$ and $n$. Therefore, rows/columns
$m$ and $n$ are merged with new $p$ value of the merged row/column as geometric
mean of the separate $p$ values of the corresponding elements.}
\label{fig:clustering}
\end{figure}

\subsubsection{Clustering Technique}

Our clustering method is as follows. We compute $p$ values for all possible
protein pairs and store them in a matrix. Then we pick the protein pair with
lowest $p$ value and choose it as the first group in the cluster. The rows
and columns for these two proteins are merged into one row and one column
[Fig.~\ref{fig:clustering}].
Probability numbers for this new group are geometric means of the two
probabilities [or arithmetic means of the $log(p)$ values]. The process is
continued repeatedly, thus adding more and more
clusters as well as making the existing ones bigger, until a threshold is reached.

\subsection{Ancient Paralogs}

Ref.~\cite{Wolfe} proposed possibility of duplication of the entire yeast
genome in some distant past and presented a list of genes that were identical
or matched closely due to this event. We check how many of the associations
derived by us were also such ancient paralogs and present them in
Table.~\ref{table:ancient}.  We find 22 such ancient paralogs among the list of
top 2833 pairs (.7\%). Therefore, these are the ancient paralogs that
maintained their functions over time.
\begin{table}
\begin{tabular}{|l|l|l|}
\hline
Protein 1 & Protein 2& Index\\
\hline
MYO3 & MYO5 & 1 \\
GIC1 & GIC2 & 72 \\
TIF4632 & TIF4631 & 145\\
NUP100 & NUP116 & 476\\
HSC82 & HSP82 & 485\\
ZDS1 & ZDS2 & 564 \\
PPH21 & PPH22 & 579\\
KCC4 & GIN4 & 606\\
RFC3 & RFC4 & 634\\
CLN1 & CLN2 & 918\\
GSP2 & GSP1 & 1288\\
YPT32 & YPT31 & 1550\\
BOI1 & BOI2 & 1640\\
SEC4 & YPT7 & 1785\\
YPT53 & VPS21 & 1888\\
BMH1 & BMH2 & 1920\\
PCL7 & PCL6 & 1926\\
YGR010W & YLR328W & 2162\\
MYO4 & MYO2 & 2474\\
SAP190 & SAP185 & 2721\\
MKK1 & MKK2 & 2725\\
IMD4 & YLR432W & 2746\\
\hline
\end{tabular}
\caption{Associations derived by us which were also ancient paralogs according
to Ref.~\cite{Wolfe}. Third column in the table represents the indices for the
pairs in the list of associations sorted according to increasing probabilities.
The list is also available as a supplementary material.
}
\label{table:ancient}
\end{table}

\section{Supplemental Informations (Functional modules derived using our technique)}

\end{document}